\documentclass[twocolumn,prl,aps,superbib,tightenlines,floatfix,superscriptaddress,showpacs]{revtex4}
\usepackage[T1]{fontenc}
\usepackage[latin9]{inputenc}
\usepackage{amsmath}
\usepackage{graphicx}
\usepackage{amssymb}

\begin{document}

\title{Spontaneous Emergence of Persistent Spin Helix from Homogeneous Spin Polarization}

\author{Valeriy A. Slipko}
\affiliation{Department of Physics and Astronomy and USC
Nanocenter, University of South Carolina, Columbia, SC 29208, USA}
\affiliation{ Department of Physics and Technology, V. N. Karazin
Kharkov National University, Kharkov 61077, Ukraine }

\author{Ibrahim Savran}

\affiliation{Department of Computer Science and Engineering, University of South Carolina, Columbia, SC 29208, USA}

\author{Yuriy V. Pershin}
\email{pershin@physics.sc.edu}

\affiliation{Department of Physics and Astronomy and USC
Nanocenter, University of South Carolina, Columbia, SC 29208, USA}

\begin{abstract}
We demonstrate that a homogeneous spin polarization in one-dimensional structures of finite length in the presence of Bychkov-Rashba
spin-orbit coupling decays spontaneously toward a persistent spin helix.
The analysis of formation of spin helical state
is presented within a novel approach based on a mapping of spin drift-diffusion equations into a heat equation
for a complex field. Such a strikingly different and simple method allows generating robust spin structures whose
properties can be tuned by the strength of the spin orbit interaction and/or structure's length. We generalize our results for two-dimensional case
predicting formation of persistent spin helix in two-dimensional channels from homogeneous spin polarization.
\end{abstract}

 \maketitle

The helical wave of rotating spin orientation is referred to as the spin helix.
There is a significant interest to spin helix configurations in semiconductor materials since the electron spin relaxation of such spin configurations can be partially \cite{Pershin05a,pershin10a,Weber07a} or even completely suppressed \cite{Bernevig06a,Koralek09a}. While a partial suppression of spin relaxation in two-dimensional systems becomes possible in the presence of only Bychkov-Rashba \cite{Bychkov84a}
spin-orbit coupling (see Refs. \cite{Pershin05a,pershin10a}), the complete suppression of spin relaxation requires a specific combination of Bychkov-Rashba and Dresselhaus \cite{Dresselhaus55a} interactions as it was demonstrated in Ref. \cite{Bernevig06a}. More generally, the relaxation of the spin helix is an example of situations \cite{Kiselev00a,Sherman03a,Weng04a,Pershin04a,Pershin05a,Jiang05,Bernevig06a,Schwab06a,Weber07a,Weng08a,Koralek09a,Kleinert09a,Duckheim09a,Tokatly10a,pershin10a} when electron spin relaxation scenario deviates from the predictions of D'yakonov-Perel' theory \cite{Dyakonov72a}.

Experimentally, the spin grating technique \cite{Cameron96a} is typically used \cite{Weber07a,Koralek09a} to create spin helical configurations in semiconductors. In this method, a sample is illuminated by a pair of pump beams with orthogonal linear polarizations. The interference of such beams results in a spacial modulation of light helicity. Correspondingly, through the optical orientation effect, a modulation of spin polarization in the form of spin helix is produced. Moreover, a spin injection from a ferromagnetic material into a semiconductor can also be used to excite a spin helix \cite{Schliemann03a}. In this approach, the rotating spin polarization is caused by coherent spin precession of electrons drifting in an applied electric field. However, the present authors are not aware about any experimental studies of spin helixes excited by spin injection.

\begin{figure}[b]
 \begin{center}
\includegraphics[angle=0,width=7.0cm]{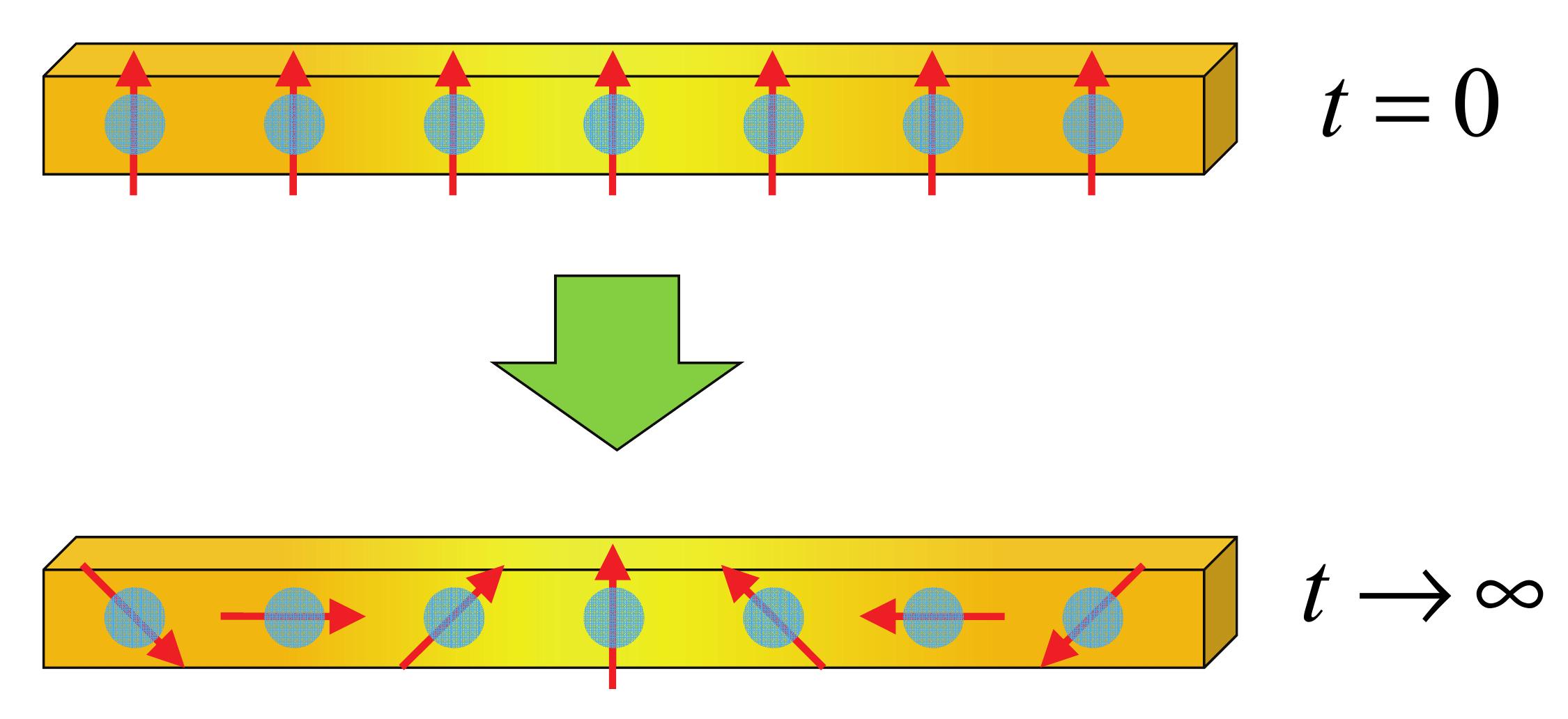}
\caption{\label{fig1}(Color online) Schematics of spontaneous transformation of homogeneous spin polarization into
persistent spin helix in a finite length system with Bychkov-Rashba spin-orbit coupling.}
 \end{center}
\end{figure}

\begin{figure*}[t]
 \begin{center}
    \centerline{
    \mbox{\includegraphics[width=8.0cm]{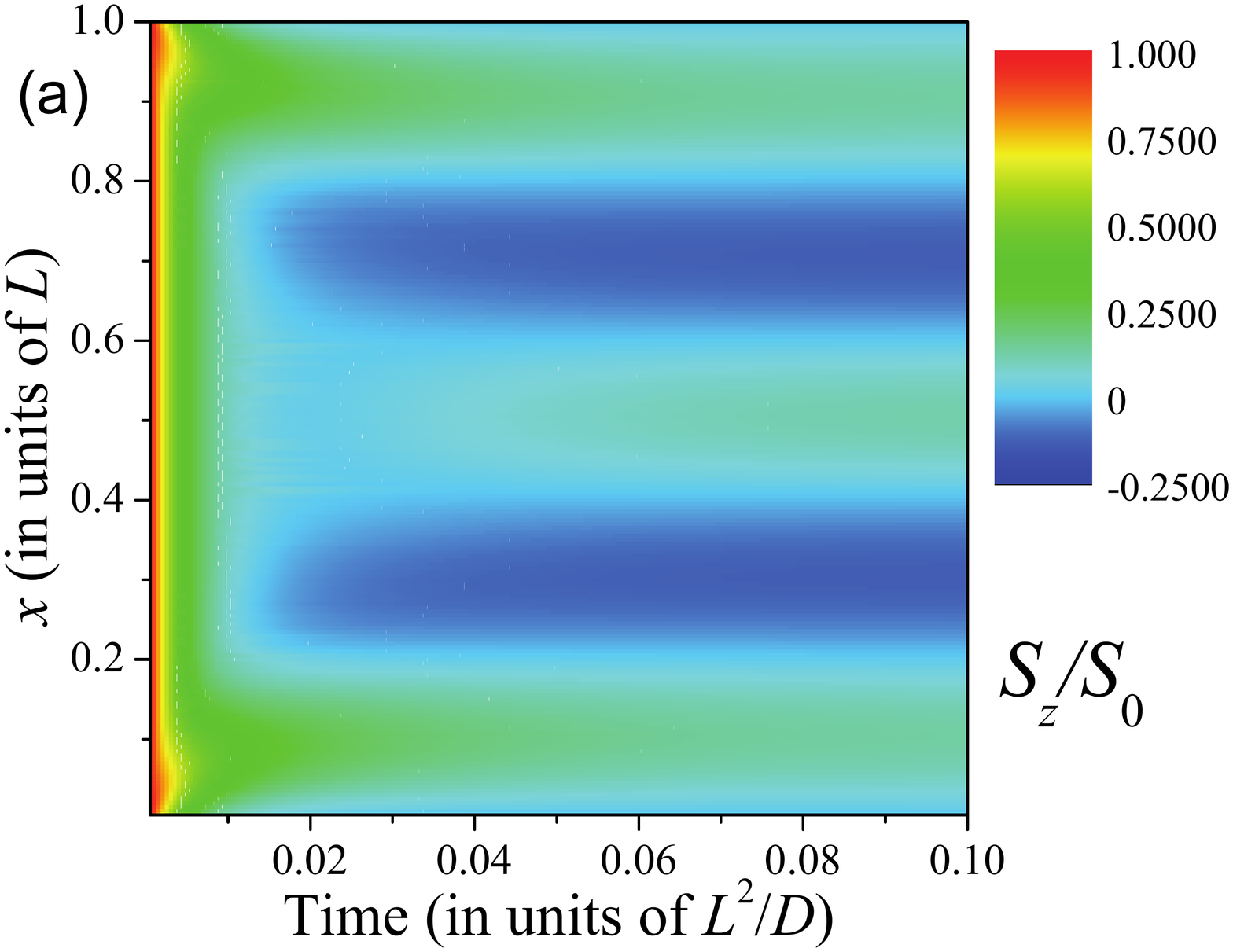}}
    \mbox{   }
    \mbox{\includegraphics[width=8.0cm]{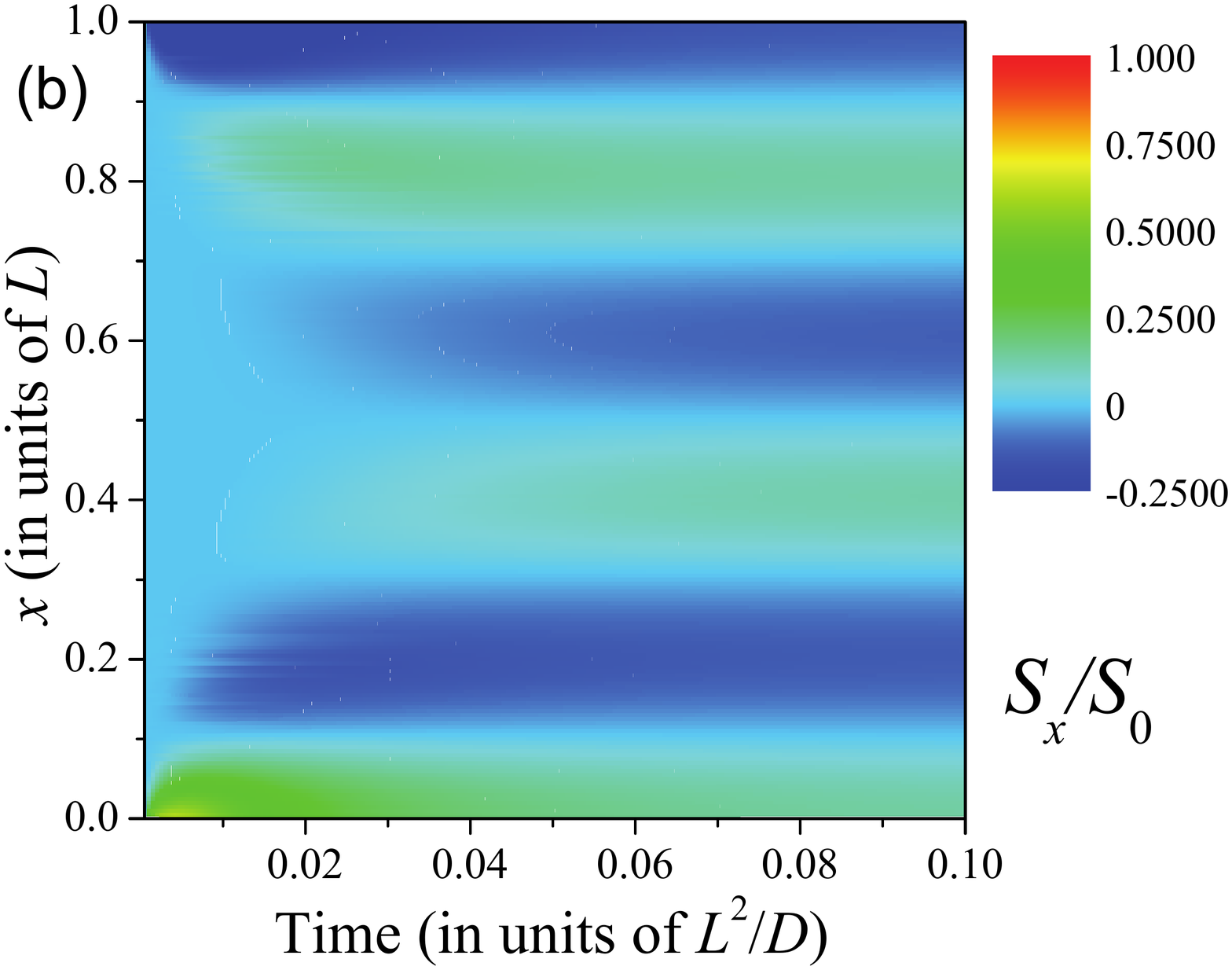}}
  }
\caption{ (Color online) Dynamics of formation of persistent spin helix from homogeneous spin polarization pointing in $z$ direction at $t=0$. These plots were obtained using Eq. (\ref{SSolution}) at $\eta L=15.45$. This value of the parameter $\eta L$ corresponds to the second local maximum of spin helix amplitude shown in Fig. \ref{fig3}.} \label{fig2}
\end{center}
\end{figure*}

In this Letter, we propose an alternative approach to induce spin helical configurations. Specifically, we demonstrate that in one-dimensional (1D) systems of finite length with Bychkov-Rashba spin-orbit coupling the spin helical configurations emerge in the process of relaxation of homogeneous spin polarization (see Fig. \ref{fig1}).  Mathematically, such a strikingly unexpected transformation of homogeneous spin polarization into the persistent spin helix occurs when we introduce boundary conditions on electron space motion to describe finite-length structures (in infinite systems the homogeneous spin polarization decays exponentially as predicted by D'yakonov-Perel' theory \cite{Dyakonov72a}). Using a novel approach that maps spin drift-diffusion equations into a heat transfer equation for a complex field we find the exact time-dependence of the spin polarization dynamics. It is interesting that the amplitude of the resulting spin helix has an oscillatory dependence on the system's length. Below, we provide an intuitive explanation of this result based on properties of solution of heat equation. Moreover, it is necessary to emphasize that our theory is generalized for the case of two-dimensional (2D) channels and can be straightforwardly verified experimentally. In particular, experimentally, the homogeneous spin polarization can be easily created using the optical orientation by circularly polarized light. Therefore, we believe that our approach would simplify tremendously the generation of long-living spin helical configurations in semiconductor structures and advance the field of spin storage in semiconductors.

Let us consider dynamics of electron spin polarization in a 1D system of a length $L$ in $x$ direction in the presence of Bychkov-Rashba spin-orbit coupling.
In one-dimensional limit, spin drift-diffusion equations \cite{pershin10a} can be written as
\begin{eqnarray}
\frac{\partial S_x}{\partial t}&=&D\Delta S_x+C\frac{\partial
S_z}{\partial x}-2\gamma S_x \label{SxEq}, \\
\frac{\partial S_y}{\partial t}&=&D\Delta S_y \label{SyEq}, \\
 \frac{\partial S_z}{\partial t}&=&D\Delta S_z-C\frac{\partial S_x}{\partial
x}-2\gamma S_z
\label{SzEq},
\end{eqnarray}
where $D=\ell^2 / \tau$ is the coefficient of diffusion, $\Delta=\partial^2 / \partial x^2$, $C=2\eta D$ is the constant describing spin rotations, $\gamma=\eta^2 D/2$ is the coefficient describing spin relaxation, $\eta=2\alpha m \hbar^{-1}$ is the spin precession angle per unit length,
$\alpha$ is the spin-orbit coupling constant, $m$ is the effective electron mass, $\ell$ is the mean free path and
$\tau$ is the momentum relaxation time. It follows from Eq. (\ref{SyEq}) that $y$ component of spin polarization, $S_y$, is not coupled to any other component of spin polarization. Consequently,
selecting $S_y(x,t=0)=0$ we can safely take out $S_y$ from our consideration.
Eqs. (\ref{SxEq},\ref{SzEq}) are complimented by standard  boundary conditions \cite{Galitski06a}

\begin{eqnarray}
\left(2D\frac{\partial S_x}{\partial x}+C S_z\right)_\Gamma=0,~
\left(2D\frac{\partial S_z}{\partial x}-C S_x\right)_\Gamma=0.
\label{BC}
\end{eqnarray}
Here, $\Gamma=[x=0,x=L]$. Mathematically, the boundary conditions (\ref{BC}) are so-called third-type boundary conditions.
This specific form of boundary conditions conserves the spin polarization of electrons that scatter from the sample edges.
We assume that at the initial moment of time the spin polarization is homogeneous and points in $z$ direction, that is
\begin{eqnarray}
S_x(x,t=0)=0,~ S_z(x,t=0)=S_0. \label{IC}
\end{eqnarray}

Let us introduce a complex polarization $S=S_x+iS_z$. It is straightforward to show
that Eqs. (\ref{SxEq},\ref{SzEq}) and boundary conditions (\ref{BC})
can be rewritten in a more compact form using $S$:
\begin{eqnarray}
\frac{\partial S}{\partial t}=D\frac{\partial^2 S}{\partial
x^2}-iC\frac{\partial S}{\partial x}-2\gamma S,
\label{ComplexDiffusionEq}
 \\
 \left(2D\frac{\partial S}{\partial x}-i C S\right)_\Gamma=0.
 \label{ComplexBC}
\end{eqnarray}
Defining a complex field $u(x,t)$ by the relation
\begin{eqnarray}
u(x,t)=e^{-i\eta x}S(x,t) \label{u},
\end{eqnarray}
we find that Eq. (\ref{ComplexDiffusionEq}) transforms into the heat equation
\begin{equation}
\frac{\partial u}{\partial t}=D\frac{\partial^2 u}{\partial x^2}, \label{uEq}
\end{equation}
supplemented by
Neumann (or second-type) boundary conditions
\begin{equation}
 \left(\frac{\partial u}{\partial x}\right)_\Gamma=0\label{uBC}.
\end{equation}
Moreover, it is worth noticing that the initial conditions for $u(x,t)$ are related to the initial conditions for $S$ as
\begin{equation}
u(x,t=0)=e^{-i\eta x}S(x,t=0). \label{InCondHeatEq}
\end{equation}
Consequently, the initially homogeneous spin polarization in $z$ direction (Eq. (\ref{IC})) corresponds to a spatially modulated complex field
\begin{equation}
u(x,t=0)=S_0 \sin\left(\eta x \right)+i S_0 \cos\left(\eta x \right). \label{uIC}
\end{equation}

The solution of Eq. (\ref{uEq}) with the boundary conditions (\ref{uBC}) and initial condition (\ref{uIC}) was obtained by the method of separation of variables. It can
be presented in the form
\begin{eqnarray}
\frac{S(x,t)}{S_0}=i\frac{\sin(\eta L/2)}{\eta L/2}e^{i\eta ( x-L/2)}+ \nonumber \, \, \, \, \, \, \, \, \, \, \, \, \\
 2\eta L e^{i\eta x}\sum_{n=1}^{+\infty}\frac{1-(-1)^n
e^{-i\eta L}}{(\eta L)^2-(\pi n)^2}e^{-\frac{\pi^2 n^2 Dt}{L^2}}\cos\left(\frac{\pi n
x}{L}\right) . \label{SSolution}
\end{eqnarray}
This is our main analytical result describing dynamics of spin polarization in 1D finite-length structures. Note that $S_x$ and $S_z$ components of spin polarization are given by real and imaginary parts of Eq. (\ref{SSolution}), respectively. The first term in the right-hand side of Eq. (\ref{SSolution}) describes the persistent profile of spin polarization (in the form of spin helix) emerging  at long times. Concerning the second term in the right-hand side of Eq. (\ref{SSolution}), it governs the dynamics of transformation of the initially homogeneous spin polarization into the persistent spin helix. Fig. \ref{fig2} demonstrates dynamics of $S_z$ and $S_x$ components of spin polarization given by Eq. (\ref{SSolution}). It is clearly seen that the initially homogeneous spin polarization in $z$ directions transforms into the persistent spin helix with an (infinitely) long lifetime.

\begin{figure}[t]
 \begin{center}
\includegraphics[angle=0,width=8.0cm]{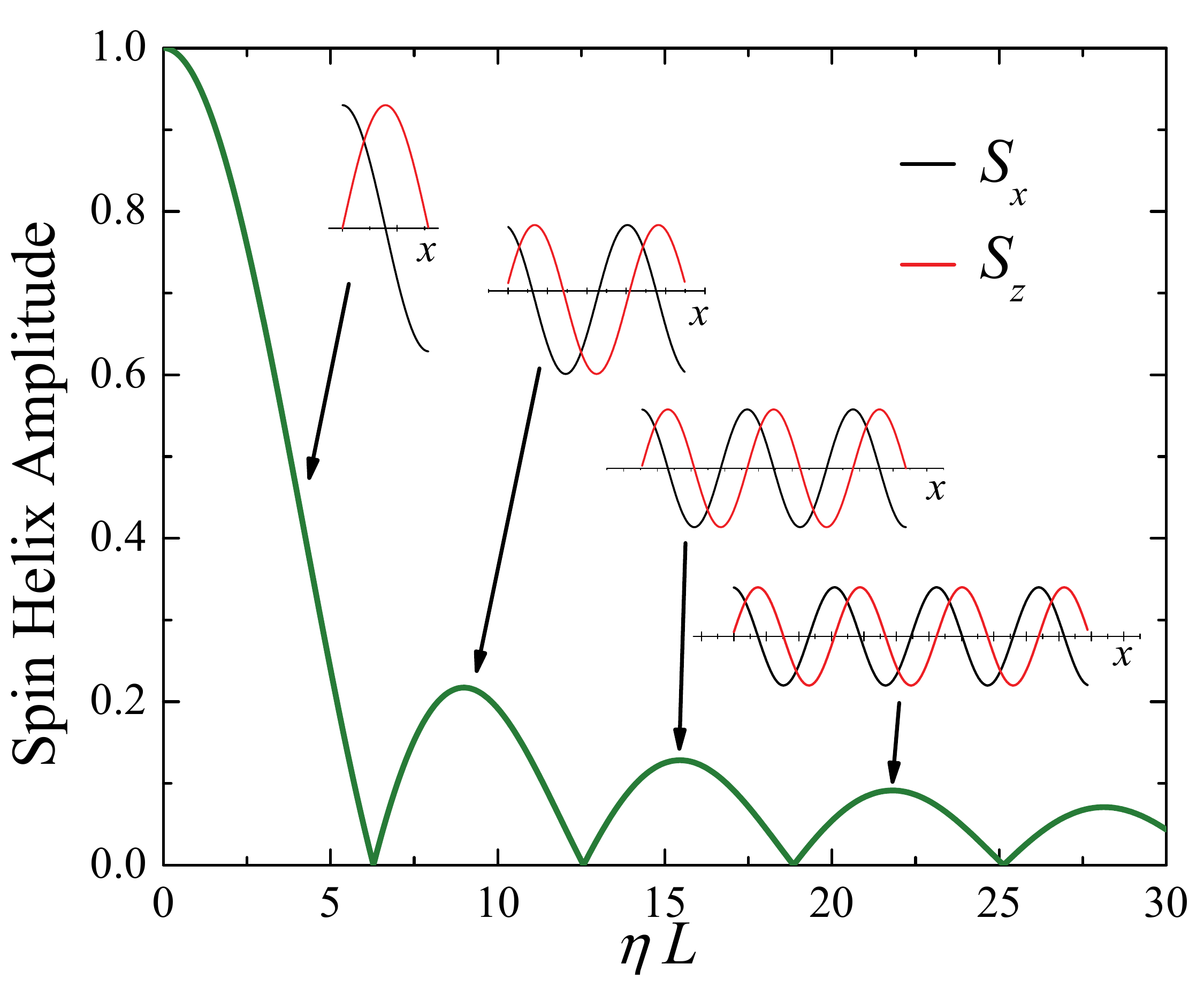}
\caption{(Color online) Normalized amplitude of the persistent spin helix as a function of $\eta L$. Insets show schematically distributions of $S_x$ and $S_z$ at several specific values of $\eta L$ as indicated by arrows. Positions of minima and maxima points of the amplitude are discussed in the text.} \label{fig3}
 \end{center}
\end{figure}

Explicitly, in the long time limit, the spin polarization is given by
\begin{eqnarray}
S_x(x,t=+\infty)=-S_0\frac{\sin(\eta L/2)}{\eta L/2} \sin (\eta (
x- L/ 2)) \label{SxInfty}, \\
S_z(x,t=+\infty)=S_0\frac{\sin(\eta L/2)}{\eta L/2} \cos (\eta (
x- L/ 2)) \label{SzInfty}.
\end{eqnarray}
In these equations the factor $\sin(\eta L/2)/(\eta L/2)$ defines reduction of the spin helix amplitude with the respect to the initial amplitude of homogeneous spin polarization $S_0$. We plot this function in Fig. \ref{fig3}. It is interesting that the  spin helix amplitude is an oscillating function of the parameter $\eta L$ and takes zero values when $\eta L=2\pi n$ where $n$ is a positive integer. The positions of local maxima can be found numerically. In particular, positions of four local maxima shown in Fig. \ref{fig3} are 8.987, 15.450, 21.808, 28.132.

The heat equation is the best starting point to understand the oscillatory dependence of spin helix amplitude on $\eta L$ depicted in Fig. \ref{fig3}. Accordingly to  Eq. (\ref{InCondHeatEq}), the initially homogeneous initial condition (Eq. (\ref{IC}) for spin diffusion equations transforms into a modulated initial condition for the heat equation. As the solution of heat equation in the given context represents simply the process of temperature equilibration along the system, an integer number of modulation periods results in zero average "temperature" and, correspondingly in zero spin helix amplitude. Moreover, we would like to mention that the spin helix formation process is described by a series of exponentially decaying terms whose time constants are given by $\tau_n=L^2/ (\pi^2 n^2 D)$. The longest of these times $\tau_1=L^2/ (\pi^2 D)$ provides the time scale of the transformation process. The dependence of spin process on $L$ is intuitively clear as electrons should "feel" the system's length before the transformation ends. It's also interesting that such a time can be longer or shorter then the relaxation time of homogeneous spin polarization $\tau_h=1/(D\eta^2)$. In particular,
$\tau_1/\tau_h=( \eta L/\pi )^2$ meaning that  $\tau_1<\tau_h$ when
$ \eta L<\pi$, the times are the same when $\eta L=\pi$, and $\tau_1>\tau_h$ when $ \eta L>\pi$ (see also Fig. \ref{fig3}).

\begin{figure}[tb]
 \begin{center}
\includegraphics[angle=0,width=8.0cm]{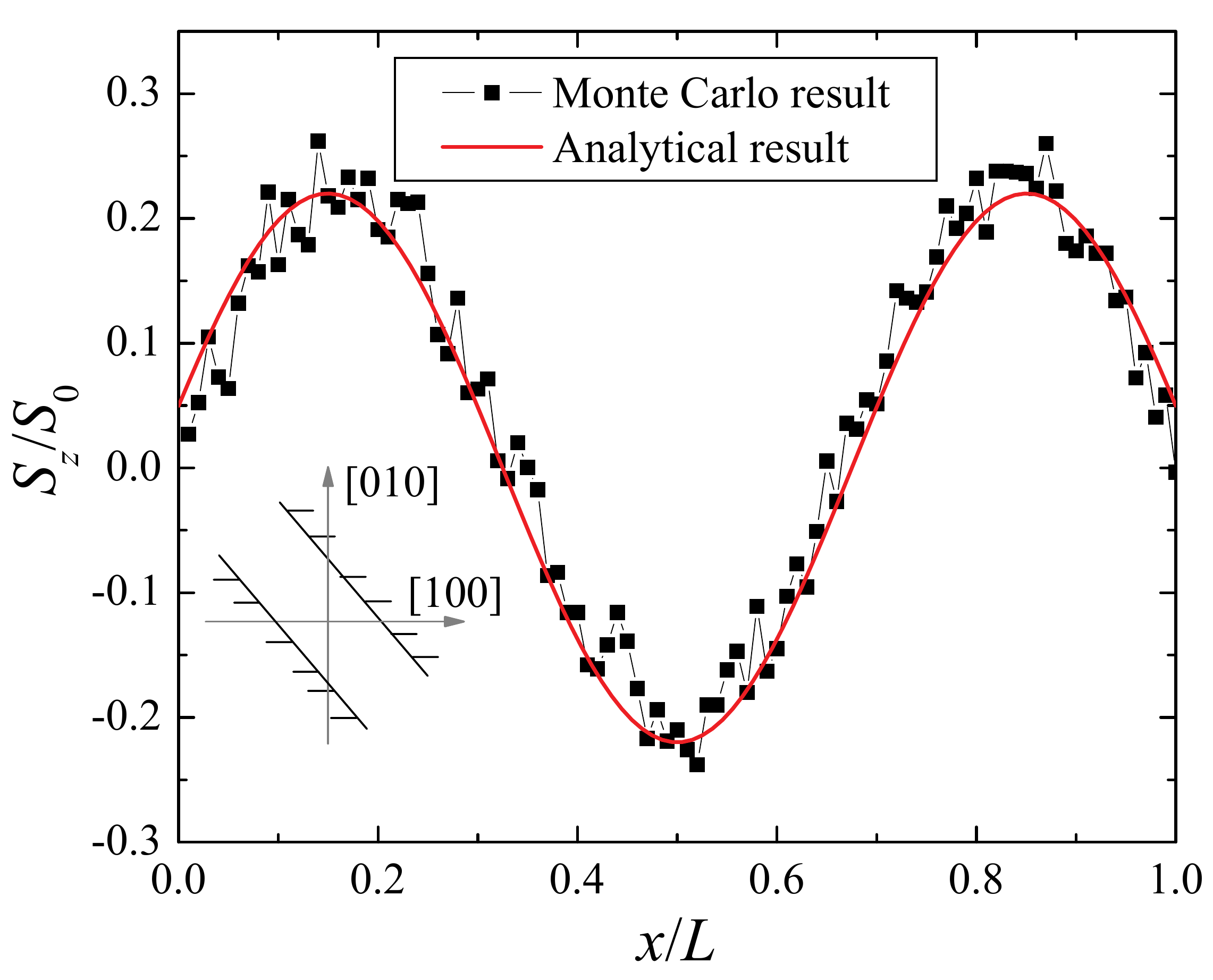}
\caption{(Color online) Long-time distribution of $S_z$  at $\eta L=8.987$ found employing Monte Carlo simulation approach. The analytical curve
is obtained using Eq. (\ref{SzInfty}). The Monte Carlo simulation was performed for $10^5$ electrons in GaAs structure of $1.7\mu$m length. This plot obtained using the parameter values $\tau=0.1$ps, $l=10$nm, $\alpha=3\cdot 10^{-12}$eV m. Inset: orientation of 2D channel for two-dimensional spin helix excitation experiments.} \label{fig4}
 \end{center}
\end{figure}

In order to obtain an additional insight on spin relaxation of the radial spin helix, we have performed extensive Monte Carlo
simulations employing an approach described in Refs. \cite{Kiselev00a} and \cite{Saikin05a}. This Monte Carlo simulation method uses a semiclassical description of electron space motion and quantum-mechanical description of spin dynamics (the later is based on the Bychkov-Rashba coupling term). All specific details of the Monte Carlo simulations program can be found in
the references cited above and will not be repeated here. A spin conservation condition was used for electrons scattering from system boundaries. Generally, all obtained Monte Carlo simulation results are in perfect quantitative
agreement with our analytical predictions thus confirming the mechanism of formation of persistent spin helix from homogeneous spin polarization. A comparison of selected analytical and numerical curves is given in Fig. \ref{fig4}.

The results reported in this paper can be readily generalized for the persistent spin helix in two dimensions \cite{Bernevig06a}. Indeed, it can be easily seen that in the case of equal strength of Bychkov-Rashba and Dresselhaus spin-orbit interactions, $\alpha=\beta$ (where $\beta$ is  the Dresselhaus spin-orbit coupling constant), the equations of spin diffusion in 2D \cite{Bernevig06a} take the general form of Eqs. (\ref{SxEq}-\ref{SzEq}). Therefore, introducing appropriate boundary conditions, namely, reducing the system into a 2D channel in [-110] direction (see the inset in Fig. \ref{fig4}), we obtain the situation completely equivalent to that in 1D from the point of view of spin dynamics. Taking into account recent experimental demonstration of  persistent spin helix \cite{Koralek09a} the emergence of persistent spin helix from homogeneous spin polarization can be straightforwardly detected. Finally, we would like to note that the amplitude of persistent spin helix can be increased by a repetitive excitation of homogeneous polarization by a train of laser pulses.

In summary, we have demonstrated that persistent spin helix forms in the process of relaxation of homogeneous spin polarization in finite length systems. This observation can be used as a different technique for creating spin helical structures in semiconductors. The solution of spin drift-diffusion equations describing formation of persistent helix was derived analytically and numerically using Monte Carlo simulation approach. The results obtained in both ways are in perfect agreement. It is interesting that the persistent helix amplitude demonstrates an oscillatory dependence on the system length and strength of spin orbit interaction. Therefore, the control of spin helix characteristics is achievable via appropriate choice of the above mentioned parameters. This suggested technique facilitates generation of spin helical states and can be used in both one- and two-dimensional geometries.

I. S. acknowledges PhD scholarship from the Republic of Turkey Ministry of National Education, Grant No: MEB1416.

\bibliography{spin}
\end{document}